\documentclass[%
 aip,
 sd,%
 amsmath,amssymb,
 reprint,%
]{revtex4-1}

\usepackage{graphicx}
\usepackage{dcolumn}
\usepackage{bm}
\usepackage{amsmath}
\usepackage[multiple]{footmisc}

\draft 

\begin{document}


\title{Gate-tunable supercurrent and multiple Andreev reflections in a superconductor-topological insulator nanoribbon-superconductor hybrid device} 



\author{Luis~A.~Jauregui}
\thanks{These authors contributed equally}
\thanks{Current address: Department of Physics, Harvard University, Cambridge, MA 02138, USA}

\affiliation{School of Electrical and Computer Engineering and Birck Nanotechnology Center, Purdue University, West Lafayette, IN 47907, USA}
\author{Morteza~Kayyalha}
\thanks{These authors contributed equally}
\affiliation{School of Electrical and Computer Engineering and Birck Nanotechnology Center, Purdue University, West Lafayette, IN 47907, USA}

\author{Aleksandr~Kazakov}
\affiliation{Department of Physics and Astronomy, Purdue University, West Lafayette, IN 47907, USA}
\author{Ireneusz~Miotkowski}
\affiliation{Department of Physics and Astronomy, Purdue University, West Lafayette, IN 47907, USA}
\author{Leonid~P.~Rokhinson}
\affiliation{Department of Physics and Astronomy, Purdue University, West Lafayette, IN 47907, USA}
\affiliation{School of Electrical and Computer Engineering and Birck Nanotechnology Center, Purdue University, West Lafayette, IN 47907, USA}
\author{Yong~P.~Chen}
\email{yongchen@purdue.edu}
\affiliation{Department of Physics and Astronomy, Purdue University, West Lafayette, IN 47907, USA}
\affiliation{School of Electrical and Computer Engineering and Birck Nanotechnology Center, Purdue University, West Lafayette, IN 47907, USA}
\affiliation{Purdue Quantum Center, Purdue University, West Lafayette, IN 47907, USA}


\date{\today}

\begin{abstract}
We report on the observation of gate-tunable proximity-induced superconductivity and multiple Andreev reflections (MAR) in a bulk-insulating BiSbTeSe$_2$ topological insulator nanoribbon (TINR) Josephson junction (JJ) with superconducting Nb contacts. We observe a gate-tunable critical current ($I_C$) for gate voltages ($V_g$) above the charge neutrality point ($V_{CNP}$), with $I_C$ as large as 430 nA. We also observe MAR peaks in the differential conductance ($dI/dV$) versus DC voltage ($V_{dc}$) across the junction corresponding to sub-harmonic peaks (at $V_{dc} = V_n = 2\Delta_{Nb}/en$, where $\Delta_{Nb}$ is the superconducting gap of the Nb contacts and $n$ is the sub-harmonic order). The sub-harmonic order, $n$, exhibits a $V_g$-dependence and reaches $n = 13$ for $V_g = 40$ V, indicating the high transparency of the Nb contacts to TINR. Our observations pave the way toward exploring the possibilities of using TINR in topologically protected devices that may host exotic physics such as Majorana fermions.
\end{abstract}


\maketitle 

Three-dimensional topological insulators (TI's) are a new class of quantum matter with an insulating bulk and conducting surface states, topologically protected against time-reversal-invariant perturbations (scattering by non-magnetic impurities such as crystalline defects and surface roughness) \cite{Hasan2010, Qi2011}. Topological superconductors (TSC's) are another important class of quantum matter and are analogous to TI's, where the superconducting gap and Majorana fermions of TSC's replace the bulk bandgap and Dirac fermion surface states of the TI, respectively \cite{Qi2011}. Controlling the Majorana modes is considered one of the important approaches for developing topologically protected quantum computers. Three-dimensional (3D) TIs in proximity to s-wave superconductors have been proposed as one of the promising platforms to realize topological superconductivity and Majorana fermions \cite{Fu2008}. In this context, it has been pointed out that TI nanowires (TINWs) possess various appealing features for such studies \cite{Cook2011,Cook2012,Ilan2014,Jauregui2015,Jauregui2016}. However, the first important step is to understand how TI nanowires, including nanoribbons (TINR's), behave in contact with superconducting leads.

Superconductor – normal – superconductor (SNS) Josephson junctions(JJs), with topological insulators as the normal material have been experimentally realized on 3D-TI's \cite{Sacepe2011,Qu2012,Veldhorst2012,Williams2012,Sochnikov2013,Oostinga2013,Finck2014,Lee2014,Yang2014,Zhang2011,Kurter2013,Wiedenmann2015,Stehno2016,Deacon2017}. However, TI materials used in many of the previous experiments have notable bulk conduction, making it challenging to distinguish from the contribution of the topological surface states. In this letter, we study S-TINR-S Josephson junctions, where S = Niobium (Nb) and the TINR's are mechanically exfoliated from bulk BiSbTeSe$_2$ (BSTS) TI crystals. Our BSTS is among the most bulk-insulating TI's with surface states dominated conduction, and chemical potential located close to the surface state Dirac point in the bulk bandgap \cite{Xu2014,Xu2016}. Therefore, our study enables us to investigate the proximity effects and induced superconductivity in such ``intrinsic'' (bulk-insulating) and gate-tunable TINR's with both electron (n) and hole (p) dominated surface transport. Moreover, we are able to investigate the transparency of our superconducting contacts to TINR both in n- and p- dominated transport regimes through the observation of multiple Andreev reflections (MAR).

High-quality single crystals of BiSbTeSe$_2$ (BSTS) were grown by the Bridgman technique as described elsewhere \cite{Xu2014,Xu2016}. Devices fabricated on the exfoliated flakes from these crystals exhibit surface dominated conduction with ambipolar field effect, half-integer quantum hall effect, and $\pi$-Berry phase \cite{Xu2014,Xu2016}. We obtain BSTS nanoribbons using a standard mechanical exfoliation technique and transferred them onto a 500-$\mu$m thick highly doped Si substrate (used as the back gate) covered with 300-nm SiO$_2$ on top. We locate BSTS nanoribbons, which are randomly dispersed on the substrate, by an optical microscope. An atomic force microscope (AFM) image of a representative JJ is shown in Fig 1a. Multiple electrodes, with electrode separation $L <$ 100 nm between the adjacent electrodes, are defined by e-beam lithography for each TINR. We then deposit 30-nm thick Nb contacts by a DC sputtering system. A short ($\sim$ 5 sec) in situ Ar ion milling prior to the metal deposition is used to remove any residues left from the lithography step and native oxides on the TINR surface. Our results presented here are taken from a TINR sample with a thickness of $\sim$ 20 nm, width of $\sim$ 250 nm, and electrode separation of $\sim$ 60 nm.

Fig. 1b depicts $R$ vs. the back-gate voltage ($V_g$) at $T$ = 10 K (above the critical temperature of our deposited superconductor, $T_C^{Nb} \sim$ 6.5 K). The charge neutrality-point voltage ($V_{CNP}$) is $\sim$ 4 V for this device. The electron- and hole-dominated regimes can be easily observed in Fig. 1b as we tune $V_g$ away from the $V_{CNP}$. Using BCS theory, we estimate the $T$ = 0 K superconducting gap as $\Delta_{Nb}$ = 1.76$k_B T_C^{Nb} \sim$ 975 $\mu$eV.

\begin{figure}
\centering\includegraphics[width=0.9\columnwidth]{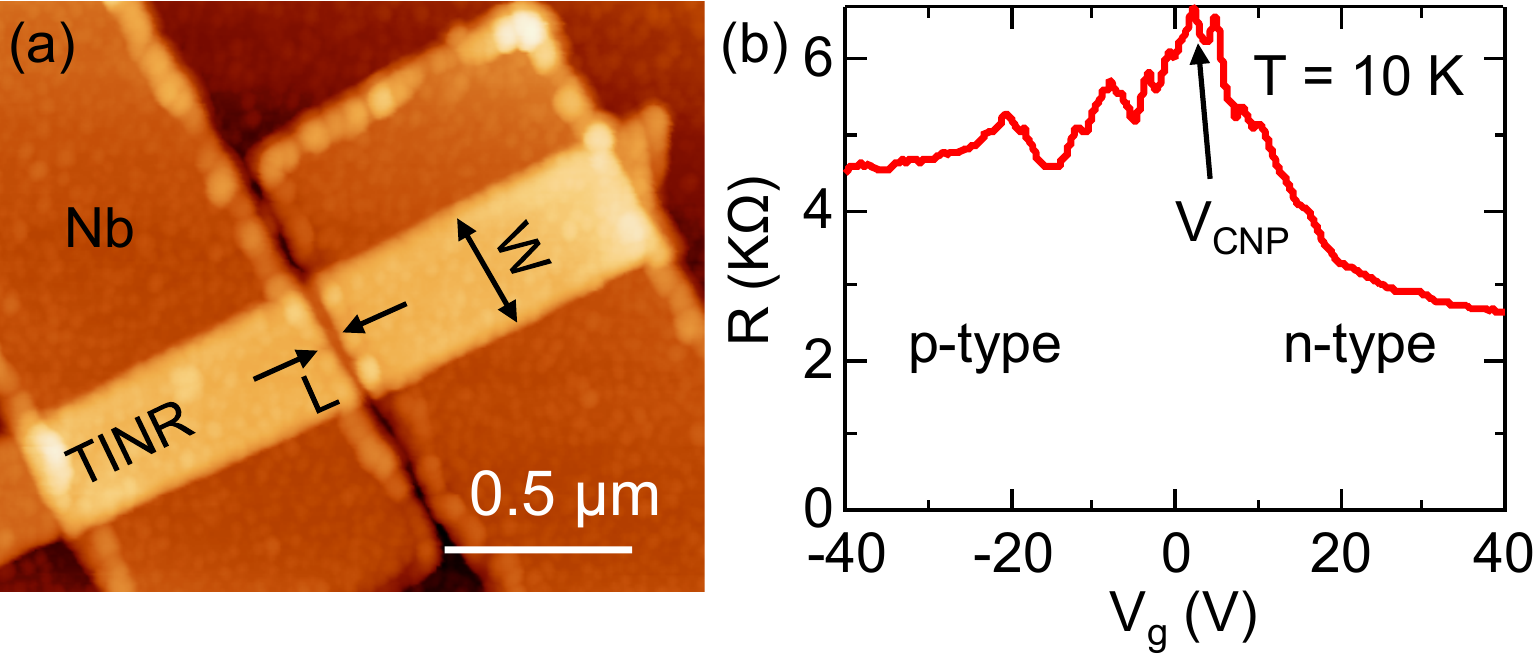}
\vspace{-0.1in}
\caption{\label{fig:epsart}
(a) Atomic force microscope (AFM) image of a 250-nm wide and 20-nm thick TINR multi-terminal device with Nb electrodes (electrode separation L $\sim$ 60 nm). (b) Two-terminal resistance ($R$) vs. the back-gate voltage ($V_g$), measured at $T$ = 10 K, above the critical temperature ($T_C^{Nb}$) of the Nb electrodes.}
\label{fig:Fig. 1}
\vspace{-0.2in} 
\end{figure}

When the sample is cooled down below $T_C^{Nb}$, the electronic transport in the junction is strongly affected by the superconducting proximity effect. The evidences of this effect manifest themselves as the flow of a supercurrent in the junction and the appearance of multiple Andreev reflections (MAR) \cite{Tinkham2004,Xiang2006}. Fig. 2a shows the colormap of the differential resistance ($dV/dI$) vs. $V_g$ and $I_{dc}$ at $T$ = 30 mK. The DC voltage vs. current ($V_{dc}$ vs. $I_{dc}$) characteristic of the junction at $T$ = 30 mK for a few different $V_g$'s is also presented in Fig. 2b. As we increase $I_{dc}$ from zero, the junction is in its superconducting state and its resistance is zero. However, once $I_{dc}$ is increased above a critical value ($I_C$, marked by an arrow in Fig 2b), the junction transitions from the superconducting state to a normal state with a non-zero resistance. The junction critical current, $I_C$, is highlighted by a white curve in Fig. 2a. First, we observe that $I_C$ is gate tunable, with larger $I_C$ for $V_g > V_{CNP}$. However, when $V_g$ is tuned near the charge neutrality point ($V_{CNP} \sim$ 4 V), $I_C$ decreases and eventually saturates for more negative $V_g$'s as previously observed in Bi$_2$Se$_3$ flakes \cite{Cho2013} and graphene  \cite{BenShalom2015,Calado2015}. One possible explanation for the saturation of $I_C$ for $V_g$ below the $V_{CNP}$ is that the Nb electrodes electron-dope the underlying material (TINR). Therefore, when $V_g < V_{CNP}$, a p-n junction is formed in the TINR. This p-n junction can weaken and eventually break the induced superconductivity as was shown in graphene \cite{Choi2013}. Another plausible explanation may be the poor injection of the holes into TINR's by Nb, as will be demonstrated from the low transparency of the contacts for $V_g < V_{CNP}$ from our analysis of MAR's (Fig. 3). The inset of Fig. 2b shows the dependence of $I_C$ on the Fermi momentum ($k_F$), where 
$k_F = \sqrt{4\pi C_{ox}(V_g-V_{CNP})/e}$ and $C_{ox}$ is the parallel plate capacitance per unit area of a 300-nm SiO$_2$ ($\sim$ 12 nF/cm$^2$). For $k_F > 0.4$ nm$^{-1}$, we observe $I_C$ varies linearly with $k_F$, as experimentally demonstrated in ballistic graphene Josephson junctions \cite{Mizuno2013}. We also observe the junction critical temperature ($T_C$, the temperature below which the junction resistance goes to zero and supercurrent starts to flow in the junction) changes with $V_g$ from $T_C$ = 1.6 K for $V_g$ = 40 V to $T_C$ = 0.7 K for $V_g$ = 10 V. Using BCS theory, we extract the induced superconducting gap ($\Delta$) in the TINR as $\Delta = 1.76k_BT_C$ = 242 $\mu$eV and 106 $\mu$eV for $V_g$ = 40 V and $V_g$ = 10 V, respectively. We note that the resistance ($dV/dI$) of the junction does not change as we increase $V_{dc}$ above $\Delta_{Nb}/e$ ($\sim 975 \mu$V) and even slightly beyond $2\Delta_{Nb}/e$ as will be discussed later. As a result, the normal resistance ($R_N$) in our junctions is obtained at $V_{dc}$ slightly above $\Delta_{Nb}/e$. We obtain $I_CR_N \sim$ 304 $\mu$V and 266 $\mu$V for $V_g$ = 40 V and 10 V, respectively.

\begin{figure}
\centering\includegraphics[width=0.9\columnwidth]{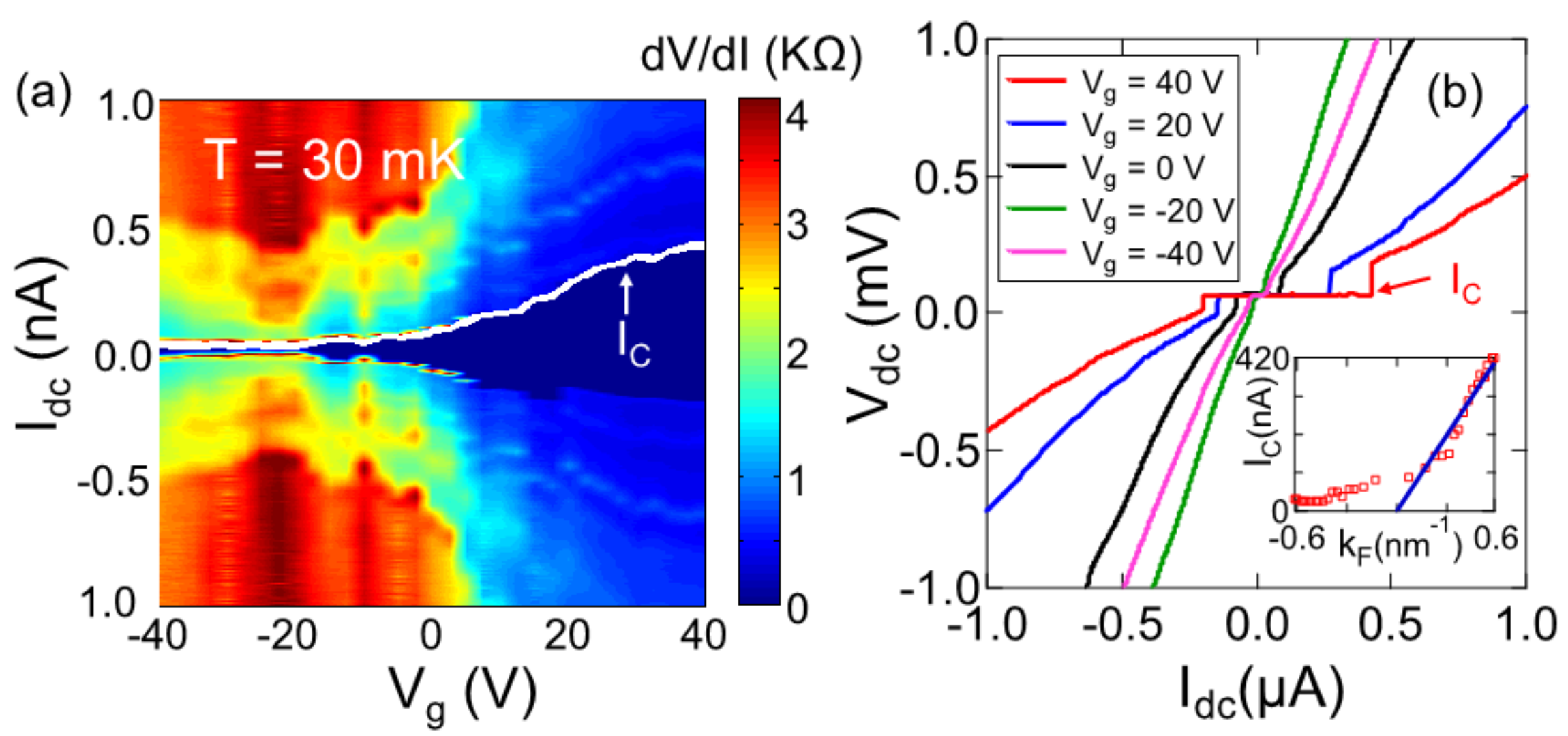}
\vspace{-0.1in}
\caption{\label{fig:epsart}
(a) Color map of $dV/dI$ vs. $V_g$ and bias current $I_{dc}$ for $T$ = 30 mK. Critical current ($I_C$) is represented by a white trace on the colormap. (b) DC Voltage ($V_{dc}$) vs. DC current ($I_{dc}$) characteristic of the device for different $V_g$'s at $T$ = 30 mK. Inset: $I_C$ vs. $k_F$ (Fermi momentum). Blue curve is a linear fit for $k_F >$ 0.4 nm$^{-1}$. Data in (a) and (b) were measured with sweeping $I_{dc}$ from $-1\mu A$ to $1\mu A$.}
\label{fig:Fig. 2}
\vspace{-0.2in} 
\end{figure}

Fig. 3a displays $dI/dV$ vs. $V_{dc}$ for $V_g$ = 40 V at $T$ = 30 mK. Several peaks (within the Nb superconducting gap) in $dI/dV$ are observed at $V_{dc} = V_n = 2\Delta_{Nb}/en$ (where $n = 2, 3, 4, 5, 6, 9,$ and 13) as marked by the arrows in Fig. 3a. These dI/dV peaks are consistent with MAR \cite{Tinkham2004}.
We note that these peaks are symmetric around $V_{dc}$ = 0 V and thus below we focus only on the positive peaks. No feature in dI/dV vs. $V_{dc}$ is identified for $n = 1$ and $R_N$ is achieved for $V > \Delta_{Nb}/e$ instead of $V>2\Delta_{Nb}/e$. The absence of the first ($n = 1$) MAR peak has been noted in some SNS junctions \cite{Wiedenmann2015,Xiang2006} and may be related to the presence of mid-gap zero-energy states as described elsewhere \cite{Badiane2011,SanJose2013}. From the linear fit of dI/dV peaks vs. $1/n$, we obtain $\Delta_{Nb} \sim 975 \mu eV$, which is in excellent agreement with the $\Delta_{Nb}$ obtained from the BCS theory and $T_C^{Nb} \sim$ 6.5 K. Moreover, the observed dI/dV peaks are reproducible and independent of $V_{dc}$ sweep direction. While we do not observe any dI/dV peaks corresponding to $n = 7$ and 8, higher-order peaks ($n = 9$ and 13) are present, a feature that needs further investigation. The observation of the high-order MAR peaks is an indication of high transparency of contacts in our junction.

Fig. 3b depicts the differential conductance (dI/dV, normalized by 1/$R_N$) vs. (positive) $V_{dc}$ for T = 30 mK at three different $V_g$'s. First, we observe that the position of the dI/dV peaks remains relatively constant with $V_g$, in contrast to the oscillatory behavior of dI/dV peaks around a resonant level in a quantum dot \cite{Buitelaar2003,JarilloHerrero2006}. This suggests the absence of localized states in our TINR devices. The high-order dI/dV peaks observed for $V_g > V_{CNP}$ further indicate that the contacts are highly transparent. However, for $V_g < V_{CNP}$, the amplitude of the dI/dV peaks decreases with more negative $V_g$, e.g. with vanishing peak amplitudes for $n = 3, 4, 5, 6,$ and 9 at $V_g$ = -40 V (see Fig. 3c). It has been previously reported that in JJs \cite{Du2008,Kjergaard2017}, the MAR peak amplitude depends on the ratio between $\xi$ (superconducting coherence length in the channel) \cite{Du2008} and L (channel length), with larger amplitudes for larger $\xi$/L. For $V_g$ = 40 V, the amplitude of the $n = 2$ peak in the normalized dI/dV is $\sim$ 2, indicating $\xi > $L, which is also corroborated with the observation of supercurrent. Fig. 3c shows details of dI/dV vs. $V_{dc}$ curves (at positive side of $V_{dc}$) for three characteristic $V_g$'s. The vanishing of dI/dV peaks for $V_g < V_{CNP}$ may be related to the pinning of the Fermi level to the electron-doped regime under the Nb electrodes and hence the formation of p-n junctions for $V_g < V_{DP}$, as has been observed in graphene JJs \cite{BenShalom2015,Calado2015}.

\begin{figure}
\centering\includegraphics[width=0.9\columnwidth]{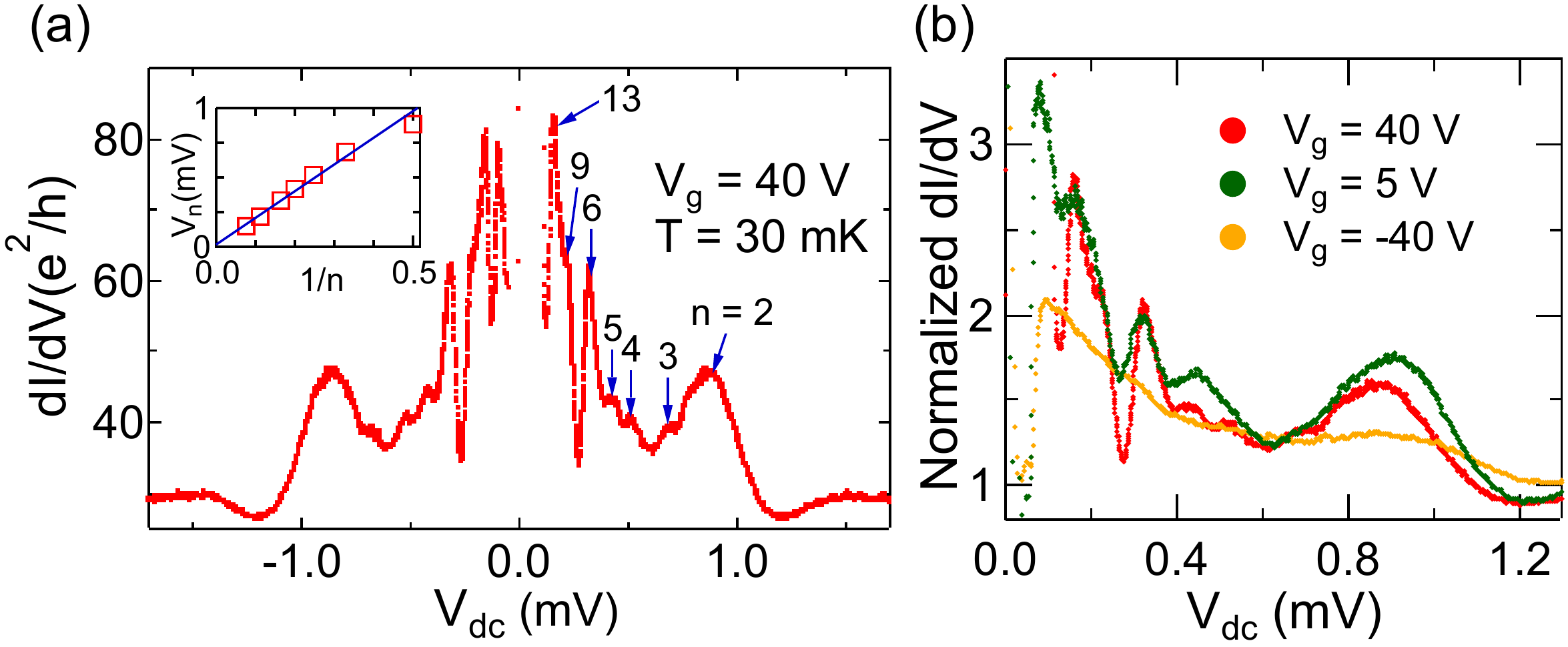}
\vspace{-0.1in}
\caption{\label{fig:epsart}
(a) Differential conductance ($dI/dV$) vs. $V_{dc}$ for $V_g$ = 40 V. Each dI/dV peak position ($V_n$, expected to be 2$\Delta_{Nb}/en$) is labeled with its index $n$, starting with $n$ = 2 for the peak near $V_{dc}$ = 900 $\mu$eV. Inset: $V_n$ vs. $1/n$. Solid line is a linear fit with a corresponding slope of $\sim$ 1.8 meV, which agrees with the 2$\Delta_{Nb}$ calculated from the BCS theory for the observed junction critical temperature $T_C \sim 6.5 K$. (b) $dI/dV$ normalized by $1/R_N$ vs. $V_{dc}$ for three representative $V_g$'s = 40, -40 and 5 V, corresponding to n-type, p-type and near the charge neutrality point. All the measurements were performed at $T$ = 30 mK.}
\label{fig:Fig. 3}
\vspace{-0.2in} 
\end{figure}

Fig. 4a depicts the T-dependence of the dI/dV (normalized by 1/$R_N$) vs. $V_{dc}$ for $V_g$ = 40 V, exhibiting a reduction of the Nb superconducting gap with increasing T. Dashed lines are guides to the eyes corresponding to the expected T-dependence of dI/dV peak positions ($V_n$) from BCS theory. We observe a nearly flat and featureless dI/dV vs. $V_{dc}$ for T = 6.6 K (slightly above $T_C^{Nb} \sim$ 6.5 K). We also observe that while dI/dV peaks are noticeable up to high temperatures $\sim$ 5.2 K), the amplitude of the peaks reduces with increasing T, and some of the peaks merge together at higher T (e.g. peaks for $n = 3$ and 4 merge at T = 3.5 K). Fig. 4b shows the T-dependence of $V_n$ for $n = 2, 3, 4$ and 6. Using the BCS theory to fit $V_n$ vs. T, we extract a $T_C \sim$ 6 K, in fair agreement with $T_C^{Nb} \sim$ 6.5 K. Fig. 4c displays the T-dependence of $\Delta_{Nb}$ extracted from each dI/dV peak (for $n = 2, 3, 4$ and 6), where $\Delta_{Nb} = neV_n(T)/2$, together with the fit of $\Delta_{Nb}$ vs. T obtained from the BCS theory, which is seen to describe the data well.

\begin{figure}
\centering\includegraphics[width=0.9\columnwidth]{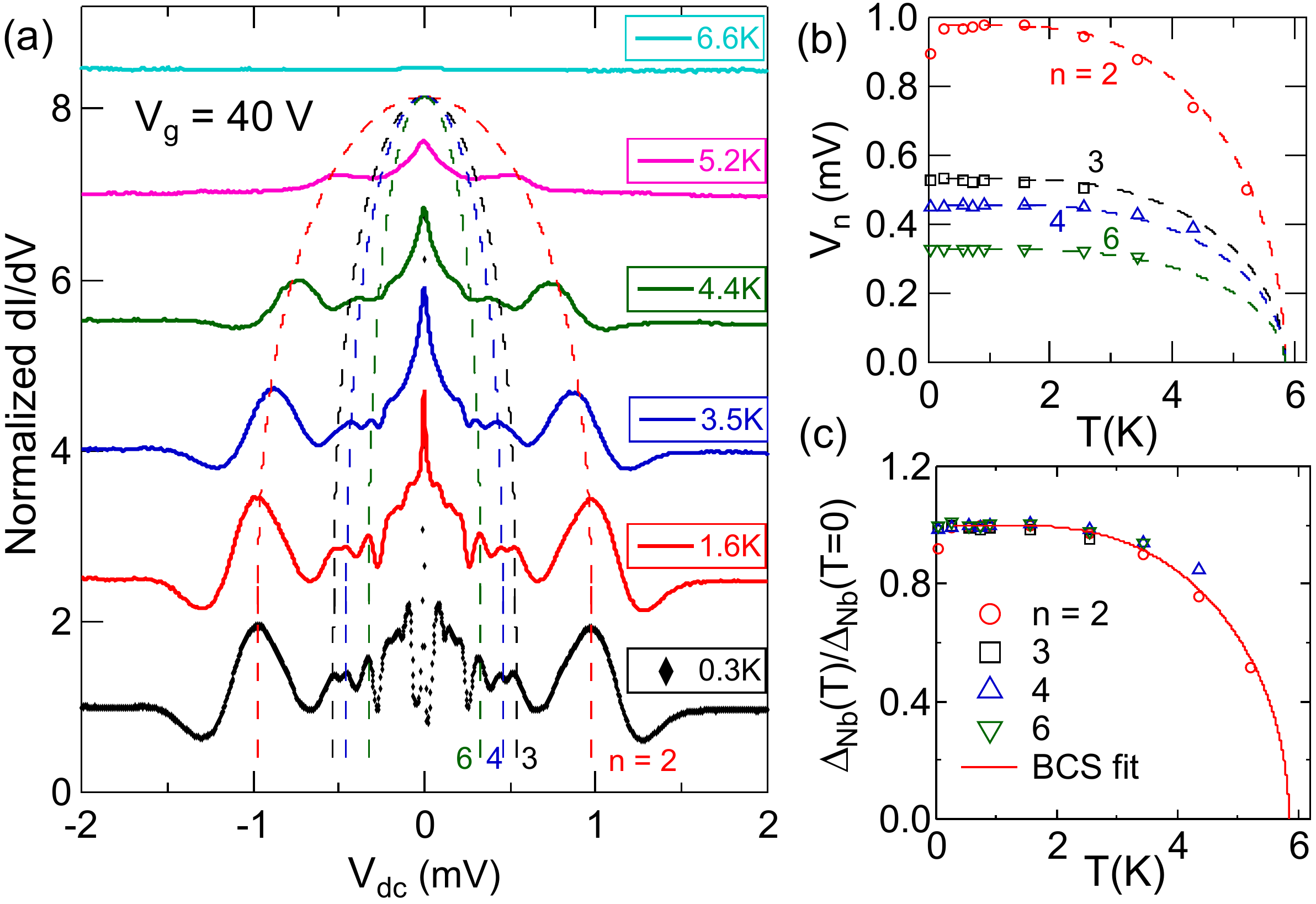}
\vspace{-0.1in}
\caption{\label{fig:epsart}
(a) Normalized $dI/dV$ vs. $V_{dc}$ for different $T$'s at $V_g$ = 40 V. Dashed lines are guides to the eyes corresponding to the expected $T$-dependence of $V_n$ from BCS theory for $n$ = 2, 3, 4 and 6. (b) $V_n$ vs. $T$ for $n$ = 2, 3, 4 and 6. Dashed lines are BCS fits. (c) Temperature dependence of normalized $\Delta_{Nb}/\Delta_{Nb}(T = 0 K)$, where $\Delta_{Nb} = enV_n(T)/2$ is obtained from different $dI/dV$ peaks corresponding to $n$ = 2, 3, 4 and 6. Solid line is a BCS-theory fit.}
\label{fig:Fig. 4}
\vspace{-0.2in} 
\end{figure}

We demonstrated Josephson junctions based on mechanically exfoliated bulk-insulating 3D topological insulator nanoribbons in proximity to superconducting Nb electrodes. We observe high-order ($n$ = 13) multiple Andreev reflections, demonstrating charge transport in the TINR channel is coherent. Furthermore, the critical current exhibits gate effects and can be gate-tuned around one order of magnitude from $\sim$ 50 nA to $\sim$ 430 nA at 30 mK. Our measurements of supercurrent in Josephson junctions based on TINRs help to better understand the nature of induced superconductivity in these junctions and pave the way toward exploration of the envisioned topologically protected devices based on superconductor-TINR-superconductor junctions.

\begin{acknowledgments}
We acknowledge support from NSF (DMR $\#$1410942). The TI material synthesis was supported by DARPA MESO program (Grant N66001-11-1-4107). L.A.J. also acknowledges support by a Purdue Center for Topological Materials fellowship. L.P.R. and A.K. acknowledge support from the U.S. Department of Energy under Award DE-SC0008630.
\end{acknowledgments}



%
%

%


\bibliography{biblography}

\end{document}